\documentstyle[psfig,times]{mn}
\begin{document}
\hsize=6truein

\renewcommand{\thefootnote}{\fnsymbol{footnote}}

\title[CMB constraints on the Cluster Baryon Fraction]
{Implications of the {\sc BOOMERANG} and {\sc MAXIMA-I} results 
for the Baryon Mass Fraction in Clusters of galaxies}

\author[]
{\parbox[]{6.in} {S. Ettori \\
\footnotesize
Institute of Astronomy, Madingley Road, Cambridge CB3 0HA \\ }} 
\maketitle
\date{May 00}

\begin{abstract}
Recent BOOMERANG and MAXIMA-I analyses of the angular power spectrum
of the Cosmic Microwave Background put more stringent constraints
on the cosmological parameters. I show that these constraints
are consistent with the observed baryon budget 
in clusters of galaxies, allowing any further contribution 
to this budget from baryons not in stars and not X-ray emitting
to be less than 14 per cent at the 90 per cent level of confidence.
\end{abstract}


\section{INTRODUCTION}

The calculations on the primordial nucleosynthesis abundance of the
light elements (e.g. D, $^3$He, $^4$He, $^7$Li) give a direct measurement 
of the value of the baryon density, $\Omega_{\rm b}$, relative
to the critical density necessary to close the Universe.
If the regions that collapse to form rich clusters maintain the same
ratio $\Omega_{\rm b}/ \Omega_{\rm c}$ as the rest of the Universe,
a measure of the cluster baryon fraction can be then used with the
estimate of $\Omega_{\rm b}$ to constrain the ``cold", clustered
component, $\Omega_{\rm c}$, of the total density parameter, 
$\Omega_{\rm tot}$.

The clusters baryon budget is composed mainly from the 
luminous baryons of the X-ray emitting gas that falls into the
cluster dark matter halo. Other contributions
come from the baryonic mass in galaxies, $f_{\rm gal}$, 
from intergalactic stars (Theuns \& Warren 1997) 
and a still hypothetical baryonic dark
matter (like MACHOs, Alcock et al. 2000; but see Freese,
Fields \& Graff 2000). 
Given the large uncertainties on the relative contribution
of the latter two sources to the baryon budget, I qualify these as
``other baryons", $f_{\rm ob}$. 
Therefore, one can put the following relation between the relative 
amount of baryons in the Universe and in clusters with total gravitating 
mass, $M_{\rm grav}$, as inferred from the equation of the hydrostatic 
equilibrium between the gas and the gravitational potential 
that does not include the self-gravity of the hot plasma (e.g. Suto, 
Sasaki \& Makino 1998):
\begin{equation}
\frac{\Omega_{\rm b} }{\Omega_{\rm c}} = \frac{M_{\rm b}}{M_{\rm grav}} =
f_{\rm gas} + f_{\rm gal} + f_{\rm ob} > f_{\rm gas},
\label{eq:fbar}
\end{equation}
where $f_{\rm gas} = M_{\rm gas}/M_{\rm grav}$, 
$f_{\rm gal} = M_{\rm gal}/M_{\rm grav}$, 
$f_{\rm ob} = M_{\rm ob}/M_{\rm grav}$. 
(Note that any dependence upon the Hubble constant is discussed
in the next section).
In the following section, I discuss this equality in more details, 
considering the corrections required to compare the cluster 
baryon fraction to the universal value.

\begin{figure}
\psfig{figure=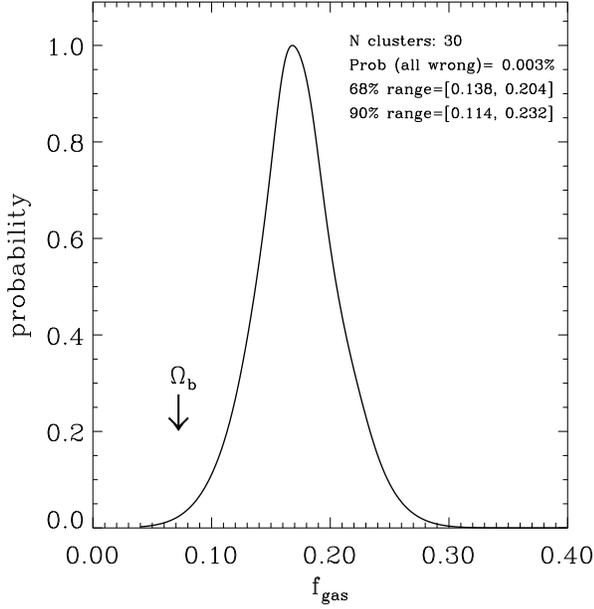,width=0.5\textwidth,angle=0}
\caption[]{Bayesian probability distribution for the gas fraction
observed in 30 high-luminosity clusters (from Ettori \& Fabian 1999). 
The distribution peaks at $f_{\rm gas} = 0.168$. 
The arrow indicates the baryon density of 0.076 $h^{-2}$ 
(Burles \& Tytler 1998) estimated through the primordial
nucleosynthesis calculations from the observed low deuterium abundance.
It has a probability of $7.2 \times 10^{-3}$ with respect to the
plotted distribution, making unlikely the agreement between the 
two baryon estimates in an Einstein-de Sitter Universe.
} \label{fig:bay} \end{figure}

When new tighter and lower constraints from nucleosynthesis were
published by Walker et al. (1991; but see discussions in recent 
years on the abundance of the light elements, e.g. Hogan 1998),
it became evident that 
just the amount of baryons visible in X-ray was enough 
to put in crisis an Einstein-de Sitter Universe with 
$\Omega_{\rm c}=1$, giving rise to the so-called 
{\it baryon catastrophe} in clusters of galaxies 
(White \& Frenk 1991, White et al. 1993,
White \& Fabian 1995, David, Jones \& Forman 1995, Evrard 1997,
Ettori \& Fabian 1999, Mohr et al. 1999; cf. Fig.~\ref{fig:bay}).

Now that independent and well-constrained measurements of the 
cosmological parameters from estimates of the angular 
power spectrum of the Cosmic Microwave Background (CMB) are
available, one can reverse the problem and, using 
the estimated values for $\Omega_{\rm b}$ and $\Omega_{\rm c}$
and the observed gas mass fraction, investigate 
the composition of the cluster baryon budget.

In particular, one can estimate the ratio
\begin{equation}
C = \frac{f_{\rm gas}}{\Omega_{\rm b}/\Omega_{\rm c}}
\end{equation}
that is expected to be close, but less than, the unity 
from eqn.~\ref{eq:fbar}.

To do this, I use  the estimates on the gas mass fraction
published in Evrard 1997 (E97, as compilation of the measurements
in White \& Fabian 1995 and David, Jones, Forman 1995), 
Ettori \& Fabian 1999 (EF99, for a sample of highly X-ray 
luminous clusters, $L_{\rm bol} > 10^{45}$ erg s$^{-1}$), 
Mohr et al. 1999 (MME99, for an X-ray flux limited sample).
I use the measurements quoted at $r_{500}$, where the mean density 
in the cluster is 500 times the background value. 
The overdensity of 500 represents a confident outer
limit where our assumptions on the intracluster gas as isothermal and
in hydrostatic equilibrium still hold.

Moreover, one can also include the stellar contribution to the baryon
clusters budget.
This is $f_{\rm gal} = M_{\rm gal}/M_{\rm grav} \approx
0.020^{+0.012}_{-0.008} h^{-1}$ (White et al. 1993, Fukugita et al. 1998).

Hereafter, I refer to $\Omega_{\rm m}$ as the {\it total} matter
density in unity of the critical density, $\rho_{\rm c} = 3H_0^2/
(8\pi G)$, where $H_0$ is the Hubble constant that we
represent as  $H_0 = 50 h_{50}$ km s$^{-1}$ Mpc$^{-1}$.
Hence, the total energy in the Universe in units of the critical value
can be written as
\begin{equation}
\Omega_{\rm tot} = (\Omega_{\rm b} + \Omega_{\rm c})
+ \Omega_{\Lambda} + \Omega_{\rm k} = \Omega_{\rm m} + \Omega_{\Lambda} 
+ \Omega_{\rm k} = 1
\end{equation}
where $\Omega_{\Lambda}$ is the constant energy density associated with
the ``vacuum" (e.g. Carroll et al. 1992) and $\Omega_{\rm k}$ is the 
energy density related to the curvature\footnote[1]{Contribution from neutrinos
is not considered because their role, i.e. their mass, is still uncertain}.
For a flat Universe, $\Omega_{\rm k}=0$ and 
$\Omega_{\rm m}+\Omega_{\Lambda}=1$.

\section{The Cosmological Constraints from BOOMERANG and MAXIMA-I}

BOOMERANG (de Bernardis et al. 2000, Lange et al. 2000) and 
MAXIMA-I (Hanany et al. 2000, Balbi et al. 2000) 
are two baloon experiments that 
looking to the CMB on scales of few tens of arcminutes 
have permitted to constrain the cosmological
models with a fit to the angular power spectrum of the 
temperature anisotropy in the detected signal.
Jaffe et al. (2001) have combined these limits with
the ones obtained from COBE/DMR (Bennett et al. 1996) on scales 
of few degrees.

I consider here the best-fit results on $\Omega_{\rm b} h_{50}^2 = 
\omega_{\rm b}$ and $\Omega_{\rm c} h_{50}^2 = \omega_{\rm c}$ quoted 
in Table~1 of Lange et al. (2001) and the values for an assumed 
$\Omega_{\rm tot} = 1$ of $\omega_{\rm b} = 0.030\pm0.004, \omega_{\rm c}
= 0.19\pm0.07, H_0 = 75\pm10$ km s$^{-1}$ Mpc$^{-1}$ in Jaffe et al. (2001).

The ratio $\omega_{\rm b}/\omega_{\rm c} = \Omega_{\rm b}/\Omega_{\rm c}$
represents the relative amount of baryons with respect
to the cold dark matter as measured in clusters of galaxies 
from the ratio of the baryonic mass to the total mass in hydrostatic 
equilibrium with the gas itself at a given radius.
Using the gas fraction estimates in E97, EF99, MME99, 
I compute the ratio $C$ that is expected to be $<1$.

Before to proceed to the measure of $C$, we have to correct  
the estimates of the observed $f_{\rm gas}$, that are generally expressed 
in terms of $h_{50} = 1$ and $\Omega_{\rm m}=1$, for changes 
in the cosmological models and known systematic errors.
These corrections are listed below and their contribution indicated
as relative error, $c_i = (f_{\rm gas}^{\rm true}-f_{\rm gas}^{\rm obs})
/ f_{\rm gas}^{\rm obs}$. 

The gas mass fraction, $f_{\rm gas}$ is given by the ratio of the gas mass and 
of the total gravitating mass within some radius, $r_{500}$ in the present 
study:
\begin{equation}
f_{\rm gas} = \frac{M_{\rm gas} (<r_{500})}{M_{\rm grav} (<r_{500})} =
\frac{4 \pi \int_0^{r_{500}} \rho_{\rm gas}(r) r^2 dr}
{ \frac{T_{\rm gas} r_{500}}{G \mu m_{\rm p}} 
\left( \frac{\partial \ln \rho_{\rm gas}}{\partial \ln r}
\right)_{r=r_{500}} },
\label{eqn:fgas}
\end{equation}
where I assume that the plasma is isothermal and in hydrostatic 
equilibrium with the gravitational potential,
$r = \theta \ d_{\rm ang}$ is the physical radius and depends upon 
the angular separation, $\theta$, and the angular diameter 
distance, $d_{\rm ang}$,
$T_{\rm gas}$ is the cluster gas temperature, $\mu$ is the mean molecular
weight in a.m.u. ($\sim 0.6$), $G$ is the gravitational constant 
and $m_{\rm p}$ is the proton mass.

Equation~\ref{eqn:fgas} depends on cosmology to define
(i) the physical radius, $r$, and (ii) the overdensity $\Delta$ here 
assumed equal to 500 for an ``$\Omega_{\rm m}=1$'' Universe. 

The {\it first cosmological correction}, $c_1$, is evaluated as follows.
Given that the surface brightness, $S(\theta)$, is the integration 
along the line of sight of the emissivity of the intracluster 
plasma due to thermal bremsstrahlung [i.e., $S(\theta) \sim \int
\rho_{\rm gas}^2 T_{\rm gas}^{0.5} dr$], 
the gas density is proportional to $d_{\rm ang}^{-0.5}$.
Combining this with the other dependence in eqn.~\ref{eqn:fgas},
\begin{equation}
f_{\rm gas} \propto \frac{d_{\rm ang}^{3-0.5}}{d_{\rm ang}} = d_{\rm ang}^{1.5}.
\end{equation}

I compute the angular diameter distance 
as a function of the cosmological parameters
$\Omega_{\rm m} + \Omega_{\Lambda} +\Omega_{\rm k} = 1$ 
(cf. eqn.~25 in Carroll, Press \& Turner 1992):
\begin{equation}
d_{\rm ang} = \frac{c}{H_0 (1+z)} \frac{S(\omega)}{|\Omega_{\rm
k}|^{1/2}}, 
\label{eq:dang}
\end{equation}
where $S(\omega)$ is sinh$(\omega)$, $\omega$, $\sin(\omega)$ for 
$\Omega_{\rm k}$ greater than, equal to and less than 0, respectively, and
\begin{equation}
\omega = |\Omega_{\rm k}|^{1/2} \int^z_0 \frac{d \zeta}{\left[ 
(1+\zeta)^2 (1+\Omega_{\rm m} \zeta) -\zeta(2+\zeta)\Omega_{\Lambda}
\right]^{1/2}}.
\end{equation}
For $\Omega_{\rm k}=0$, eqn.~\ref{eq:dang} can be written as 
\begin{equation}
d_{\rm ang} = \frac{c}{H_0 (1+z)} \times \left\{ \begin{array}{ll}
   z  & \mbox{if $\Omega_{\rm m} = 0$} \\
   \int^z_0 \frac{\Omega_{\rm m}^{-1/2} \ d \zeta }{\left[ (1+\zeta)^3 
+\Omega_{\rm m}^{-1} -1 \right]^{1/2}}  & \mbox{otherwise}
\end{array} \right.
\label{eq:dang_k0}
\end{equation}

\begin{figure*}
\hbox{
\psfig{figure=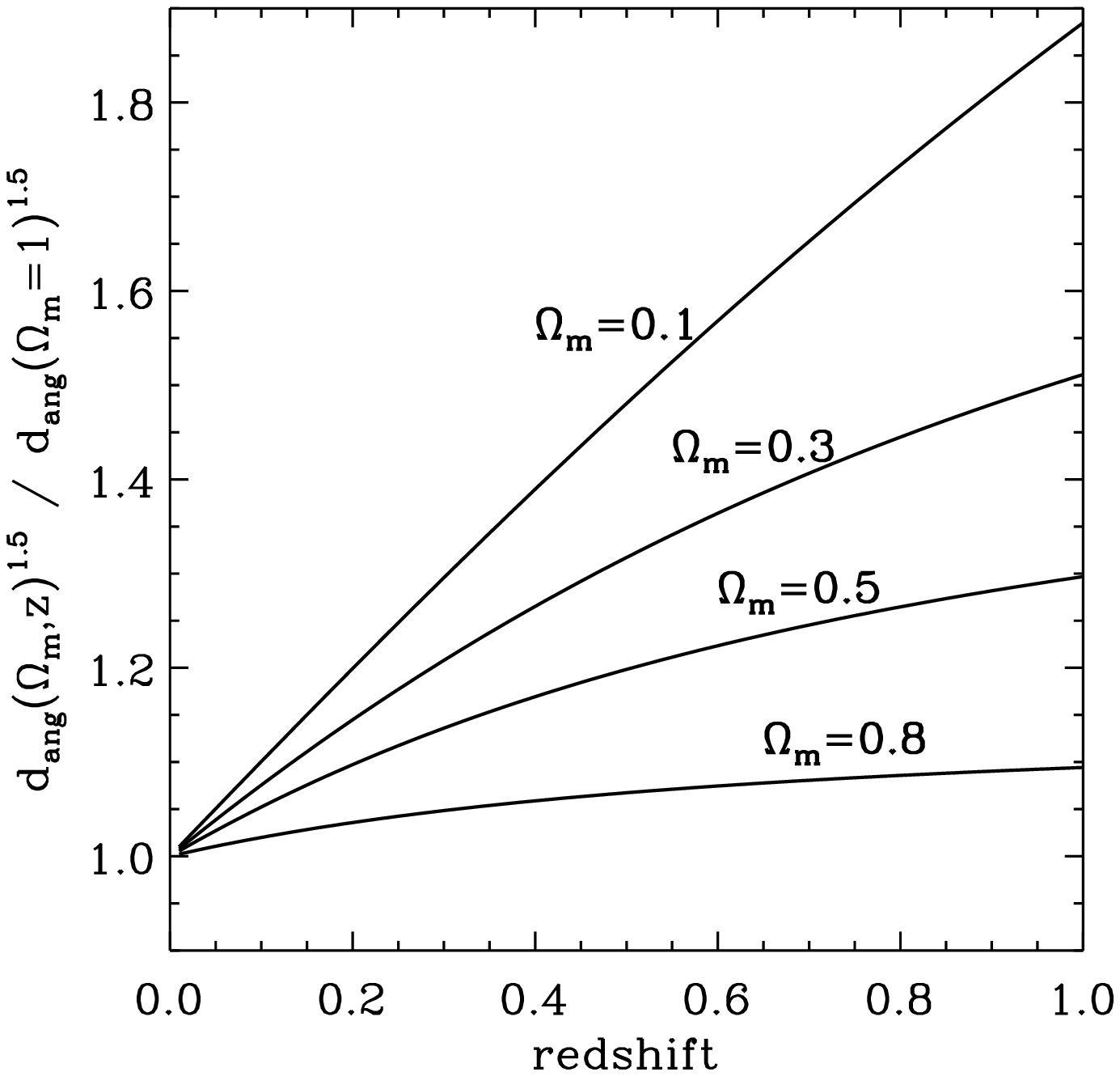,width=0.5\textwidth}
\psfig{figure=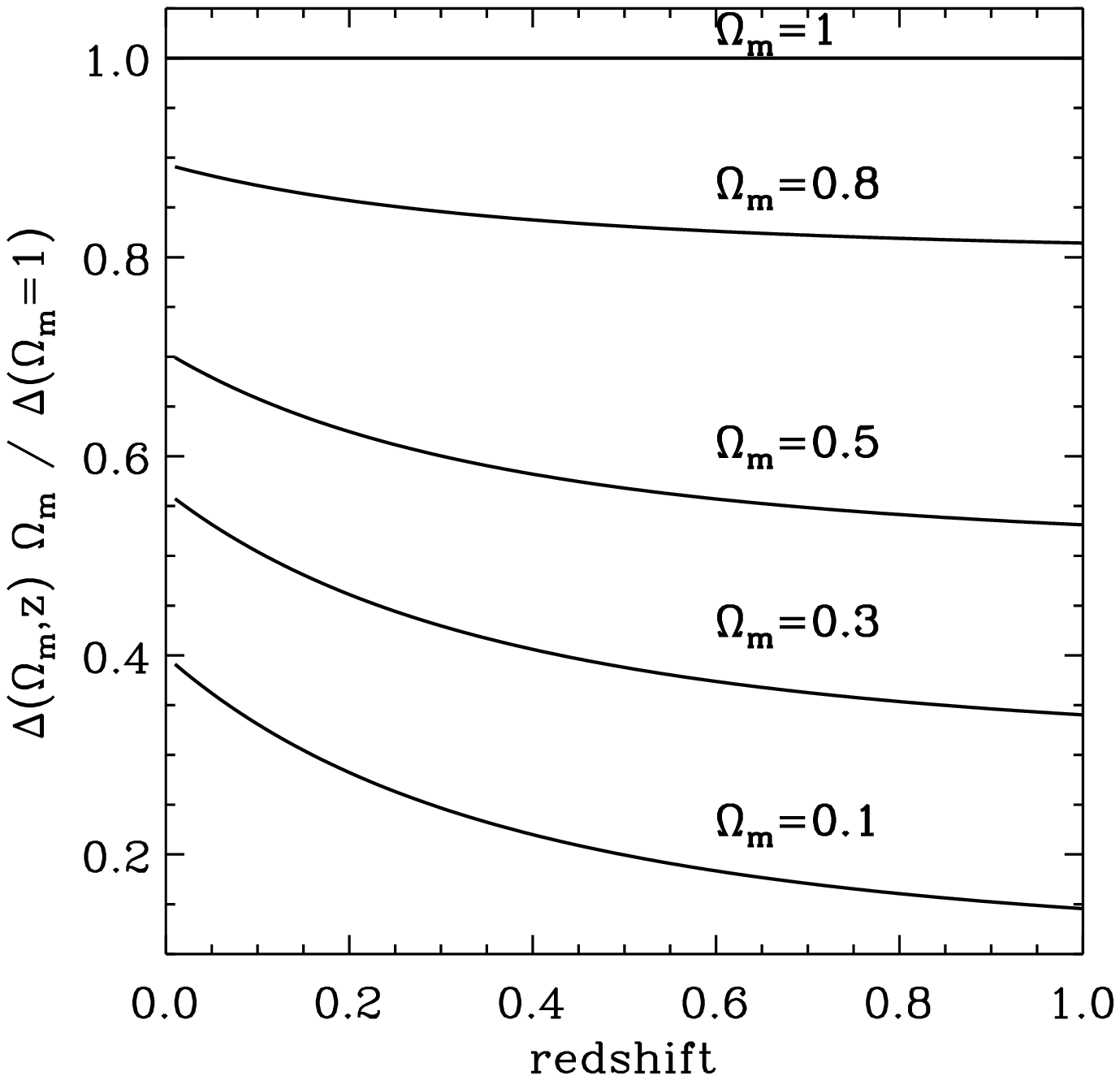,width=0.5\textwidth} }
\caption[]{(Left panel) Ratio between the dependence of the gas fraction
on the angular diameter distance 
varying $\Omega_{\rm m}$ and with $\Omega_{\rm m}=1$.
The same Hubble constant is assumed.   
(Right panel) Ratio between the overdensity, $\Delta$, 
as function of redshift and $\Omega_{\rm m}$ to the present value 
of 500 adopted in this study.
} \label{fig:cosmo} \end{figure*}

The correction, $c_1$, introduced by a low matter density Universe and 
a Hubble constant $h_{50} \sim 1.5$ lower the gas fraction by about 40
per cent ($c_1 \approx -0.40$) mainly due to the dependence upon
the Hubble constant ($f_{\rm gas} \propto h_{50}^{-1.5}$).
As shown in Fig.~\ref{fig:cosmo}, a low density Universe rises
the gas fraction measurement for nearby clusters by about 10 per cent.

The {\it second cosmological correction}, $c_2$, appears as variation 
of the radius, $r_{\Delta}$, at which the given mean overdensity of the 
total mass within the cluster, $\Delta$, with respect to the background value, 
$\Omega_{\rm m} \rho_{\rm c}$, is reached in different cosmological scenarios.
To estimate $\Delta$ in function of the cosmological parameters,
I use the relation valid for a ``$\Omega_{\rm m} +\Omega_{\Lambda}=1$" 
Universe (e.g. Kitayama \& Suto 1996, Henry 2000; see Fig.~\ref{fig:cosmo}):
\begin{equation}
\Delta(\Omega_{\rm m},z) = \Delta(\Omega_{\rm m}=1) \times
\left[ 1+0.4093 \left(\Omega_{{\rm m},z}^{-1} - 1 \right)^{0.9052} \right],
\end{equation}
where $(\Omega_{{\rm m},z}^{-1} -1) =(\Omega_{\rm m}^{-1} -1)/(1+z)^3$.

The measurement of $r_{\Delta}$ requires an observed mass profile 
$M_{\rm grav}(r)$, so that $M_{\rm grav}(<r_{\Delta})= (4/3) \pi r_{\Delta}^3 
\Delta \Omega_{\rm m} \rho_{\rm c} (1+z)^3$.
Considering that $M_{\rm grav}(r) \propto r$ in the region of interest,
\begin{equation}
r_{\Delta} = r_{500} \left( \frac{\Delta \Omega_{\rm m}}{500} \right)^{-0.5},
\end{equation}
where 500 is the mean overdensity adopted in the samples of data 
considered here for an ``$\Omega_{\rm m}=1$'' Universe.
Observations (David et al. 1995, EF99) and
hydrodynamical simulations (see E97) agree on the radial dependence
of the gas fraction near $r_{500}$, $f_{\rm gas} \propto r^{0.2}$.
Thus, the gas fraction has to be corrected by
\begin{equation}
f_{\rm gas}(r_{\Delta}) = f_{\rm gas}(r_{500}) \ 
\left( \frac{\Delta \Omega_{\rm m}}{500} \right)^{-0.1}.
\end{equation}
This correction increases the gas fraction value by about 
6 per cent or less ($c_2 \approx +0.06$).

A {\it third correction}, $c_3$, is required from the evidence 
in both simulations
and observations of baryon depletion in galaxy clusters.
The combined N-body and gas dynamics simulations
suggest that the amount of the cosmic baryons that falls
in clusters as gas component at $r_{500}$ is slightly lower than 1
($\sim 0.9$, Frenk et al. 1999; $\sim$ 0.88 in MME99), 
in agreement with the observed increase of the gas fraction with 
radius as $r^{0.2}$ near to $r_{500}$ (EF99,
David et al. 1995).
This implies that the estimated gas fraction (and the derived value
$C$) might be underestimated by about 12 per cent ($c_3 \approx
+0.12$).

A {\it fourth correction}, $c_4$, comes still from hydrodynamical simulations
of galaxy clusters that indicate an overestimate of the 
gas mass due to the neglected clumping of the plasma.
Considering that the observed thermal emission scales proportionally 
to $\rho_{\rm gas}^2$, the presence of X-ray emitting clumps
induce an overestimate on the gas fraction by about 12 per cent
respect to the ``true" value in simulated clusters, as 
shown by MME99 ($c_4 \approx -0.12$).

Other systematic errors present in the measurements of the gas fraction
and whose magnitude is still uncertain are discussed in Section~2.2.

Our analysis now requires an estimate of the ``true'' gas mass fraction 
given the four corrections listed above:
\begin{equation}
f_{\rm gas} = f_{\rm gas}^{\rm obs} \ \left( 1+\sum c_i \right),
\end{equation}
Note that just the two ones related to the cosmological models assumed
play a relevant role, because the third and fourth ones cancel 
(i.e. $c_3 + c_4 \approx 0$).

\begin{figure*}
\hbox{
\psfig{figure=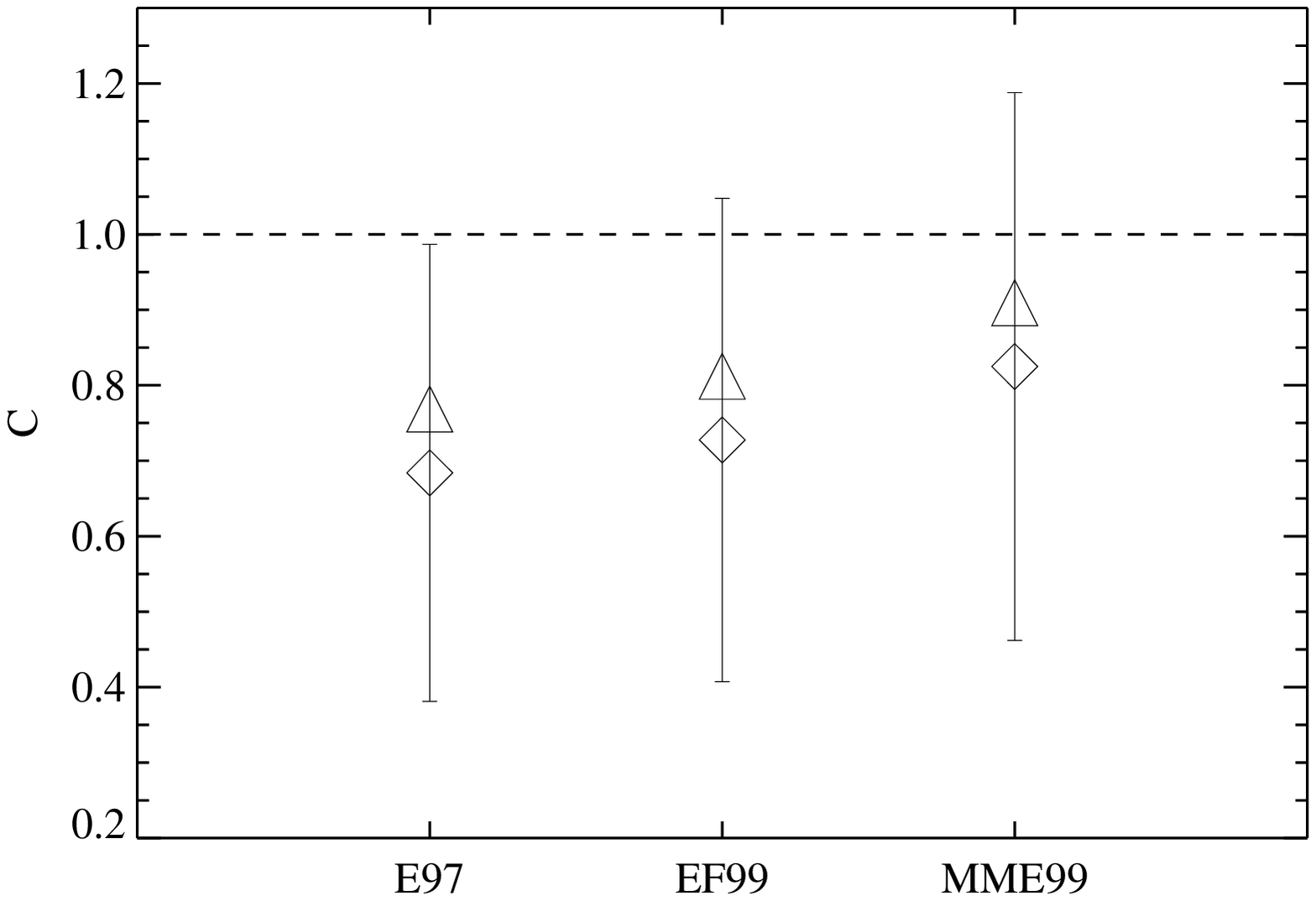,width=0.5\textwidth,angle=0}
\psfig{figure=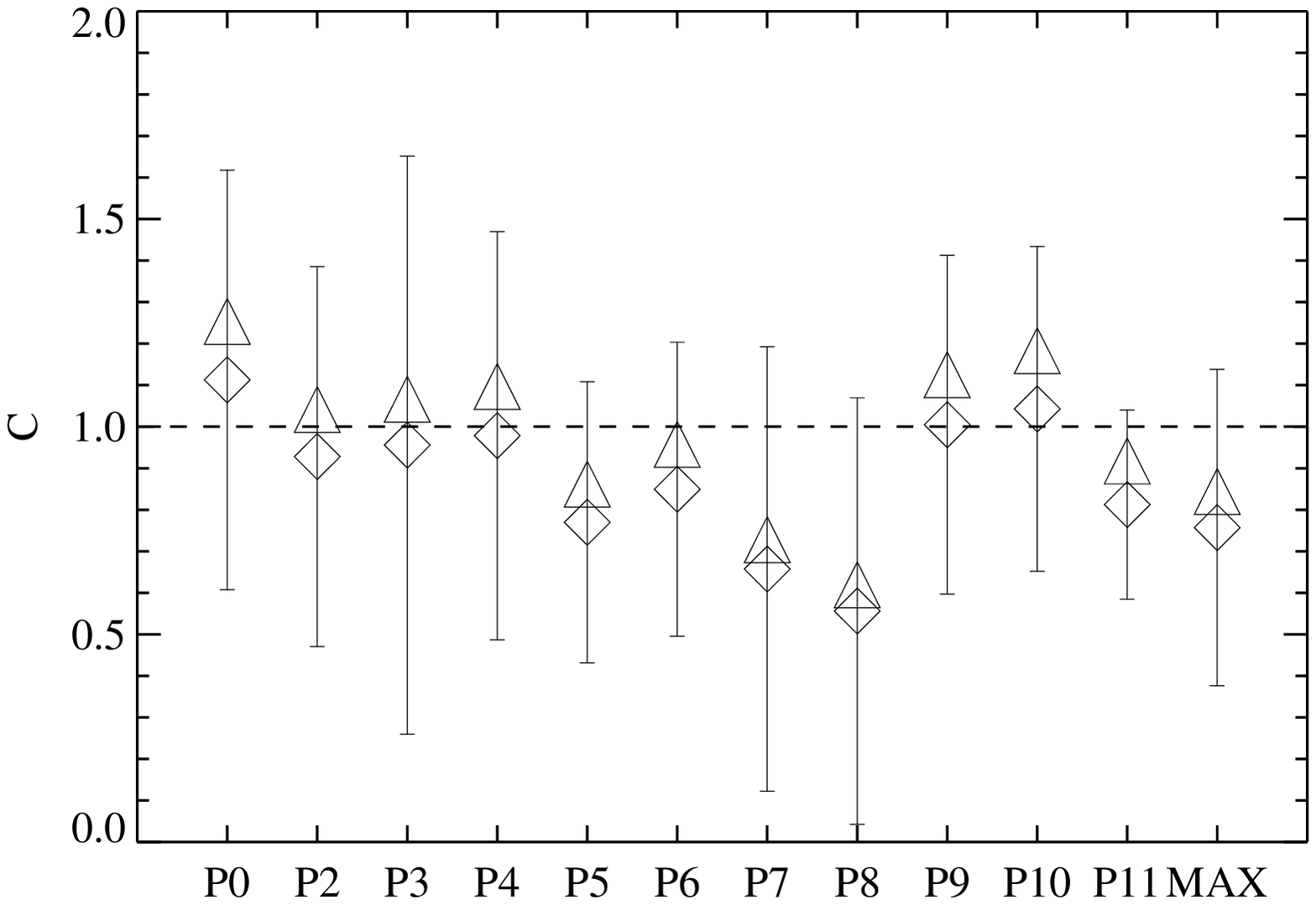,width=0.5\textwidth,angle=0}
}
\caption[]{{\it Left panel} Estimates of the ratio $C$ for gas 
fraction data in Evrard (1997, E97), Ettori \& Fabian (1999, EF99),
Mohr et al. (1999, MME99) and the joint cosmological constraints 
from BOOMERANG, MAXIMA-I and COBE/DMR in Jaffe et al. (2001).
{\it Right panel} Assuming the gas fraction value from the compilation
in Evrard (1997), I calculate $C$ for the corresponding hypotheses
$Pi$ (with $i$ from 0 to 11) in Lange et al. (2001) and for 
the best-fit values from the MAXIMA experiment. 
The {\it diamonds} indicate the central value given the only gas 
fraction, whereas the {\it triangles} include also the stellar
contribution. The error bars come from the propagation of the 
uncertainties on the considered parameters at $1 \sigma$ level and
refer to the estimates of $C$ with the gas fraction only. 
} \label{fig:boom} \end{figure*}

I calculate a weighted mean value of the distribution in the gas fraction 
measurements (after the cosmological corrections are applied and 
including the quoted errors)
for each dataset independently:
\begin{equation}
f_{\rm gas} = \left\{ \begin{array}{l}
0.108 \pm 0.005 \ {\rm (E97)} \\
0.115 \pm 0.002 \ {\rm (EF99)} \\
0.130 \pm 0.001 \ {\rm (MME99)}
\end{array}
\right.
\end{equation}
I plot the results for $C$ in Fig.~\ref{fig:boom}.

I observe that (i) $C \sim 1$ for all the sets of cosmological values
discussed in Table~1 of Lange and collaborators 
(see Fig.~\ref{fig:boom}), providing evidence that there is
now agreement between the favorite cosmological scenario and 
the observed amount of baryons in galaxy clusters,
(ii) none of these cosmological sets can be excluded
on the basis of the cluster baryon budget due to 
the large (a relative error of 40 per cent at $1 \sigma$ level)
uncertainty still presents in the estimate of $\omega_{\rm c}$.  

\subsection{Cluster baryon budget: gas, galaxies and ... then?}

From the estimate of $C$, and including the contribution 
from stars in galaxies, we are left with plausible constraints
on any additional baryonic components, like baryons in form
of dark matter or intergalactic stars:
\begin{equation}
f_{\rm ob} = \frac{\Omega_{\rm b}}{\Omega_{\rm c}} -f_{\rm gas}
-f_{\rm gal} = (1 - C - \frac{f_{\rm gal}}
{\Omega_{\rm b}/\Omega_{\rm c}}) \times \frac{\Omega_{\rm b}}{\Omega_{\rm c}}.
\end{equation}  

From the best-fit results in Jaffe et al. (2001), I obtain that
(error at $1 \sigma$ level):
\begin{equation}
f_{\rm ob} = \left\{ \begin{array}{l}
0.037 \pm 0.064 \ {\rm (E97)} \\
0.030 \pm 0.064 \ {\rm (EF99)} \\
0.014 \pm 0.064 \ {\rm (MME99)} 
\end{array}
\right.
\end{equation}
Therefore, no more than 14 per cent (90 per cent confidence level
$= 1.64 \sigma$; 23 per cent at $3 \sigma =$ 99.7 per cent 
level of confidence)
of the total matter in clusters of galaxies can be baryons not
accounted for in X-ray emitting plasma and stars in galaxies.

\subsection{Systematic uncertainties on $f_{\rm gas}$}

The assumption of an isothermal gas in hydrostatic equilibrium
with the gravitational potential provides an useful framework
to work out the contribution of the X-ray emitting plasma 
to the cluster baryon budget.
However, several aspects of the physics of the intracluster medium 
are still under investigations and can introduce some systematics 
errors on the measure of the gas mass fraction.
I list here the most significant of these, considering their
effect on the estimates of the values of $C$ and $f_{\rm ob}$.

\begin{itemize}

\item While the gas mass is well constrained from
present X-ray observations, the total mass profile still represents a
significant uncertainty in the the baryon fraction estimates.
Particularly, the shape of cluster temperature profiles is still 
defined with contradictory results 
(Markevitch 1998, Irwin et al. 1999, White 2000).
I have assumed here an isothermal gas. Now, I investigate the
variation of $f_{\rm gas}$ once a polytropic 
profile for the intracluster gas ($T_{\rm gas} \propto 
\rho_{\rm gas}^{\gamma-1}$) is considered. I model the
changes in the total gravitating mass using a $\beta-$model 
(Cavaliere \& Fusco-Femiano 1976, Ettori 2000) 
with parameters $r_{\rm c} = 0.3$Mpc, $\beta=2/3$ and
a polytropic index, $\gamma$, equal to 1.24 (Markevitch 1998). 

\item The presence of a magnetic field in the cluster plasma,
like the recent reported measurements of Faraday Rotation 
(Clarke et al. 1999, Taylor et al. 1999) 
and excess in emission in the Hard X-ray 
band (Fusco-Femiano et al. 1999) seem suggest, 
might support a non-thermal component that (i) contributes 
to the total pressure and (ii) mimics the thermal emission
by a contribution of about 10-25 per cent (Fusco-Femiano et al. 1999)
that has to be accounted for once the cluster luminosity is
recovered from the X-ray count rate. 
Assuming a proportional dependence between the non-thermal pressure,
$P_{\rm NT}$, and the gas pressure, $P_{\rm gas}$, such as 
$P_{\rm NT} = \nu P_{\rm gas}$, 
the total pressure is then $P_{\rm tot} = (1+\nu) P_{\rm gas}$ and
affects the total gravitating mass through the hydrostatic equation.
I also include the correction by a factor $\epsilon \sim \sqrt{0.80}
\sim 0.90$ of the gas density due to the non-thermal contribution
to the total observed emissivity that can be reduced .
\end{itemize}

Referring to the corrected quantities as NEW with respect to the 
previous, OLD ones, I can write:
\begin{equation}
\begin{array}{l}
M_{\rm gas, NEW} = \epsilon M_{\rm gas, OLD} \\
M_{\rm grav, NEW} = (1+\nu)  \gamma \frac{T_{\rm gas}(r)}{T_0} 
M_{\rm grav, OLD} \\
f_{\rm gas, NEW} = \frac{\epsilon}{\gamma (1+\nu)}
\frac{T_0}{T_{\rm gas}(r)} f_{\rm gas, OLD} 
\end{array}
\end{equation}

\begin{figure}
\psfig{figure=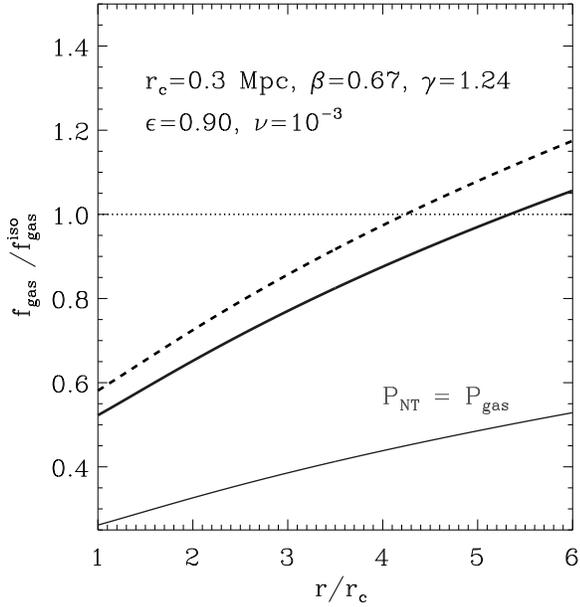,width=0.5\textwidth}
\caption{Main uncertainties on the gas fraction represented as
changes with respect to $f_{\rm gas}^{\rm iso}$ estimated from a plasma 
assumed isothermal and with no contribution from non-thermal
component. 
The dashed line represents the gas fraction corrected by the presence
of a temperature gradient with polytropic index, $\gamma=1.24$.
The solid thick line includes a correction for a non-thermal component
with the quoted $\epsilon$ and $\nu$. The thin solid line assumes
$P_{\rm tot} = 2 P_{\rm gas}$.
A magnetic field, $B$, of $\sim \mu G$ has an associated 
$P_{\rm NT}$ of about 50 K cm$^{-3} \sim 10^{-3} P_{\rm gas}$. 
Note that the presence of a temperature gradient tends 
to increase the gas fraction value at $r \sim 5r_{\rm c} \sim r_{500}$,
whereas correcting by the contamination of a non-thermal component
reduces the gas fraction.
} \label{fig:unc} \end{figure}

I plot in Fig.~\ref{fig:unc} how the gas fraction estimated 
under the isothermal assumption changes for the corrections discussed 
here.
I note that the presence of a temperature gradient reduces the
total mass and, therefore, increases the gas fraction value,
whereas a non-thermal spectral component lowers the observed thermal
contribution and, consequently, the estimate of the gas fraction
at $r_{500}$.

Then, a negative gradient in the plasma temperature profile
can reduce up to the 40 per cent the room available for any
further baryonic contribution, only partially compensated
(if below the limit of the equipartition with thermal plasma) 
by the presence of a non-thermal component.  

\section{Conclusions}

Comparing the recent cosmological constraints from measurements
of the angular power spectrum of the temperature anisotropy
in the CMB with the observed distribution of the gas mass
fraction in clusters of galaxies, I conclude that

\begin{enumerate}

\item the ratio $C = f_{\rm gas}/(\Omega_{\rm b}/\Omega_{\rm c})$
is close to 1 (Fig.~\ref{fig:boom}) for all the sets of cosmological 
values discussed in Jaffe et al. (2001) and Lange et al. (2001).
Although I can not ruled out any of these models just
from cluster baryon budget due to the large relative error
that affects $\omega_{\rm c}$, on the other hand it shows
that the most favorite cosmological scenario is consistent
with the observed baryon budget in clusters of galaxies;

\item there is evidence that only 14 per cent
(90 per cent confidence level; 
23 per cent at $3 \sigma$ level of confidence)
of the total mass can be present in galaxy clusters in the 
form of baryons neither in stars nor X-ray emitting.
It is worth to note that using the calculations in Freese et al. (2000)
and B-band luminosity and X-ray mass of the Coma Cluster in White 
et al. (1993), I estimate a contribution of MACHOs to the cluster
mass, $f_{\rm MACHO} = M_{\rm MACHO}/M_{\rm grav}$, of about
2 per cent, that is within the above constraint. 

\item the upper limit of 14 per cent 
is in the order of the observed scatter in
the gas mass distribution of about 15--20 per cent
(cf. Fig.~\ref{fig:bay}), suggesting that part of this scatter
might be due to still unobserved baryons.
It is worth to note, however, that the present accuracy of the gas
mass measurements is in the order of 10 per cent (cf. simulations in MME99)
and probably larger uncertainty is still related to the gravitating 
mass estimates (see next item). 
The combination of these uncertainties could explain the 
observed scatter in the gas mass fraction distribution. 
Data from {\it Chandra} and {\it XMM-Newton} X-ray observatories 
are in condition to reduce significantly the size of 
these uncertainties;

\item systematic errors on the gas mass fraction measurement,
like the presence of a temperature gradient in the plasma or 
a non-thermal component in the cluster emission, tends to reduce
by a value between 25 and 40 per cent the estimate of the fraction 
of the gravitating mass as ``unseen" baryons.

\end{enumerate} 

\section*{ACKNOWLEDGEMENTS} 
I am grateful to the anonymous referee for comments which
improved the presentation of this work.
I acknowledge the support of the Royal Society.

\end{document}